\documentclass[11pt]{report}

\usepackage{fancyhdr}
\usepackage{fancybox}
\usepackage{tipa}
\usepackage{framed}
\usepackage[utf8x]{inputenc}
\usepackage{epsfig, subfig, dblfloatfix}
\usepackage{graphicx}
\usepackage{anysize}
\usepackage[english]{babel}
\usepackage{caption}
\usepackage{float}
\usepackage{amsmath, amsfonts, amssymb, amsthm}
\usepackage[plainpages=false,pdfpagelabels]{hyperref}
\usepackage{abstract}
\usepackage[titletoc]{appendix}
\usepackage{listings}
\usepackage{multicol}
\usepackage{breakurl}

\usepackage{algpseudocode}
\usepackage{algorithm}
\newcommand{\B}{\ensuremath{\mathsf{B}}}

\hypersetup{
  colorlinks = true, 
  urlcolor   = blue, 
  linkcolor  = blue, 
  citecolor  = red   
}

\newcommand{\slist}{
 \begin{list}{$\bullet$}
  { \setlength{\itemsep}{0pt}
     \setlength{\parsep}{3pt}
     \setlength{\topsep}{3pt}
     \setlength{\partopsep}{0pt}
     \setlength{\leftmargin}{2em}
     \setlength{\labelwidth}{1em}
     \setlength{\labelsep}{0.5em} 
     \setlength{\listparindent}{\parindent}
     } }

\newcommand{\slisttwo}{
 \begin{list}{$\bullet$}
  { \setlength{\itemsep}{0pt}
    \setlength{\parsep}{0pt}
    \setlength{\topsep}{0pt}
    \setlength{\partopsep}{0pt}
    \setlength{\leftmargin}{2em}
    \setlength{\labelwidth}{1.5em}
    \setlength{\labelsep}{0.5em} 
    \setlength{\listparindent}{\parindent}
    } }

\newcommand{\slistend}{
  \end{list}  }

{\begin{Sbox}\begin{minipage}}%
{\end{minipage}\end{Sbox}\fbox{\TheSbox}}

\title{\textbf{C Analyzer \\A Static Program Analysis Tool for C Programs}}
\vspace{2.5cm}
\author{\bf{Project Thesis}\\
  \\
  {Submitted in partial fulfillment of requirements}\\
  \\
  {of the degree of}\\
  \\
  {\bf{MASTER OF TECHNOLOGY}}\\
  \\
  {by}\\
  \\
  {\bf{Rajendra Kumar Solanki}}\\
  \bf{Roll No. 113050074}\\
  \\
  {under supervisor:}\\
  \\
  {\bf{Prof. Krishna S. N.}}\\
  \\\\
  \includegraphics[height=3.5cm]{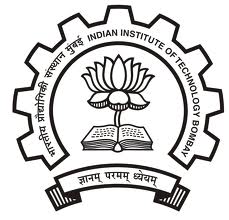}\\
  \\
  \\
  {\bf{DEPARTMENT OF COMPUTER SCIENCE AND ENGINEERING}}\\
  {\bf{INDIAN INSTITUTE OF TECHNOLOGY, BOMBAY}}\\
  }
\date{June, 2013}

\marginsize{2.5cm}{2.5cm}{1cm}{1.5cm}
\begin{document}

\pagenumbering{gobble}

\maketitle

\newpage
\vspace*{7cm}

  \begin{center}
    \Large{{To\\ \emph{Khushi} and \emph{Jatin} ...}}
  \end{center}

\pagebreak


\begin{figure}[ht]
\centering
  \includegraphics[scale=.7, angle=0]{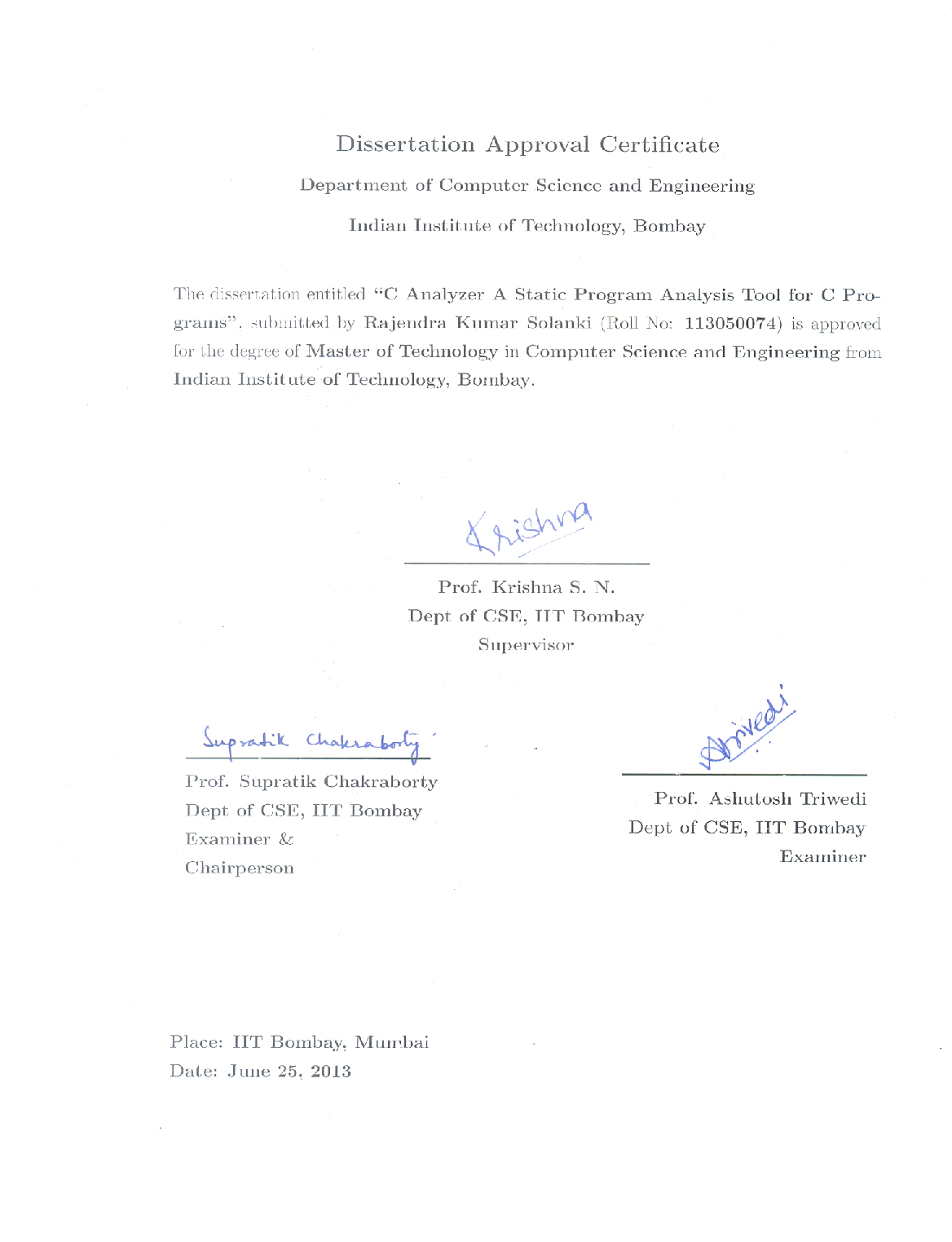}
\end{figure}

\pagebreak


\begin{figure}[ht]
\centering
  \includegraphics[scale=0.7, angle=0]{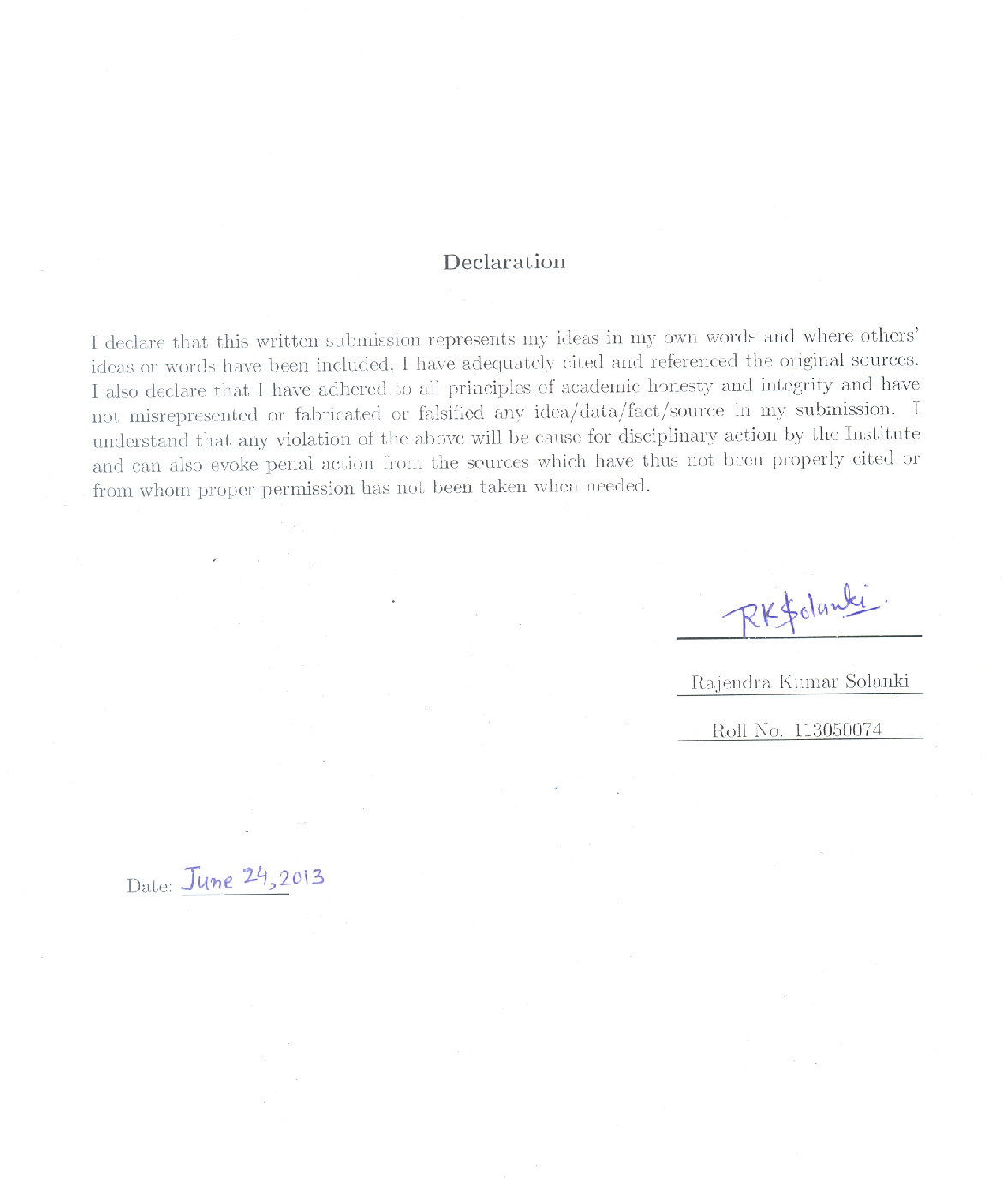}
\end{figure}

\pagebreak

\vspace*{4cm}

\begin{center}\section*{Acknowledgement}\end{center}\addcontentsline{toc}{chapter}{Acknowledgement}

\noindent I would like to thank my supervisor Prof. Krishna S. N. for this project. This would not have been possible without her, for me, to continue working in this direction after seminar in the same area. I thank Prof. Supratik Chakraborty (CFDVS) and Dr. Anup Bhattacharjee (BARC) for their queries, guidance and the progress of this work. \\

\noindent I thank Hrishikesh Karmarkar (CFDVS) and Prateek Saxena (BARC) for regular interactions on progress and challenges involved in this project over past many months. All discussions were informative and helpful in order to have better understanding of project goals. I thank Babita Sharma for her help on Clang example during initial days when we had just started working on Clang. I thank Anuj Thakkar for his support, precious time and all discussions we had during the course of this project. Anuj has been a great help on few difficult occasions to sit along with me and resolve the problem - also this has helped me to approach problems in a better way. \\

\noindent I thank my family for their unrelenting support during my academic endeavors and I am grateful to them for their support always. Finally, I thank Clang developer forum for answering my queries and all those who have helped me during this work in any possible way.

\pagebreak

\renewcommand{\abstractnamefont}{\normalfont\Large\bfseries}
\begin{abstract}
\addcontentsline{toc}{chapter}{Abstract}

\noindent In our times, when the world is increasingly getting more dependent on software programs, writing bug-free correct programs is crucial. Program verification based on formal methods can guarantee this by detecting run-time errors in safety critical systems to avoid possible adverse impact to human life and save time and money. \\

\noindent Static program analysis based on Abstract Interpretation has been in literature for quite some time. This project work tries to leverage the same for static analysis of C programs. C Analyzer is a tool developed for static analysis of C programs. This implementation of C Analyzer provides a plug-and-play domain architecture for multiple abstract domains to be used. C Analyzer supports four abstract domains - Interval, Octagon, Polyhedra and Bit Vector. We use these different domains for required precision in program verification. C Analyzer tool makes best use of LLVM's C/C++ compiler Clang's API to generate and traverse Control Flow Graph (CFG) of a given C program. This tool generates invariants in different abstract domains for statements in basic blocks of CFG during CFG traversal. Using these invariants some properties of program such as divide by zero, modulus zero, arithmetic overflow, etc. can be analyzed. We also use a source-to-source transformation tool CIL (Common Intermediate language) to transform some C constructs into simpler constructs such transforming logical operators, switch statement and conditional operator into if-else ladder and transform do-while and for loops into while loop. \\

\noindent Using C Analyzer, C program constructs such as declarations, assignments, binary operations (arithmetic, relational, bitwise shift, etc.), conditions (if-else), loops (while, do while, for loop), nested conditions and nested loops can be analyzed. Currently this tool doesn't support arrays, structures, unions, pointers and function calls.

\end{abstract}

\pagenumbering{roman}

\tableofcontents\addcontentsline{toc}{chapter}{Contents}
\listoffigures\addcontentsline{toc}{chapter}{List of Figures}

\pagestyle{fancy}



\chapter{Introduction}
\pagenumbering{arabic}
\label{sec:Introduction}


\section{Correctness of Programs}
It is often observed that writing a correct software program is difficult and to ensure that a program (or large code base) is really bug-free is even more difficult. In a world that is increasingly getting more dependent on software programs, correctness of programs is very crucial. Especially, if these programs are written for safety-critical real time applications or applications that have significant impact to human life - correctness assumes first preference. Otherwise, results can be very costly, e.g. Ariane rocket launcher failure minutes after its launch in 1996 due to an arithmetic overflow, Humburg-Altona railway switch crash in 1995 due to stack overflow, Intel chip's floating point division bug leading to millions of dollars of loss, \cite{Intro-bugs} etc. \\

\noindent \emph{Conventional testing}, that depends on program execution and a set of test cases for certain input and expected output, is costly and not exhaustive enough to ensure correctness of programs. Therefore, we need \emph{formal methods} to provide mathematically sound techniques to guarantee full coverage of all program behaviors.


\section{Static Program Analysis}
Static Program Analysis is aimed to infer program properties without executing the program. Results for static analysis are computed from given source code itself. Program verification using static analysis tries to prove absence of run-time errors in a program, without need to execute it and it checks that any operation of the program never produces errors like divide by zero, overflow, etc. \\

\noindent During static program analysis, a program is translated into a system of equations or constraints over a partial order of program properties. The solution to this system represents correct information about the particular program property being analyzed for \cite{Intro-SPA}. \\

\noindent \emph{What properties are we interested in?} Using static program analysis, we can answer some questions about a program being analyzed, such as can there be a divide by zero, or array index out of bound, what values a program variable can take, uninitialized variable being used in some arithmetic or relational operation, possible arithmetic overflow of program variables, assertion verification for safety properties, etc. \\

\noindent Our static program analysis is based on the theory of \emph{Abstract Interpretation} for proving correctness of the programs using collecting semantics of the programs.


\section{Abstract Interpretation}
Abstract Interpretation is a general theory behind approximation of program semantics. This theory is based on two main concepts: the correspondence between concrete and abstract semantics through Galois connection/insertion and the feasibility of a fixed point computation of abstract semantics using the combination of widening operators (to get faster convergence) and narrowing operators (to improve the precision of resulting analysis) \cite{WIDNARO}. \\

\noindent By generating invariants for every statement and expression in the program, we compute an approximate analysis of the program using Abstract Interpretation. In chapter \ref{sec:LiteratureSurvey}, section \ref{sec:AbstractInterpretation} this has been described in further detail. \\

\noindent Using Abstract Interpretation, static analyzers can be developed, that can automatically find properties of run-time behaviors of a program. These analyzers are sound by construction. Some spurious results (\emph{false positives}) can be produced but no scenario, and hence no bug, is left out. There is always a possibility of trade-off between precision and accuracy in such analysis. Some precision may be lost but approximation will give false positives only on safe side. \\

\noindent Abstract Interpretation was formalized by Patrick Cousot and Radhia Cousot in 1977 \cite{Intro-AI}. \\

\noindent Abstract Interpretation is: 
  \begin{itemize}
    \item \emph{sound}: due to abstract semantics being a super set of concrete semantics, covering all possible program behaviors.
    \item \emph{incomplete}: due to lack of precision some false positives may be signaled. This is the price we pay for functional correctness.
  \end{itemize}

\noindent \emph{Applications}: Abstract Interpretation finds applications in areas of specification for static program analyzers used for high-performance compilers, static analysis of programs for safety critical systems, etc. Refer \cite{Intro-ExAI} for examples of Abstract Interpretation based static analysis.


\section{Problem Statement}

\begin{center}
\emph{To build a tool for Static Program Analysis using theory of Abstract Interpretation.} \\
\end{center}

\noindent The approach that we have adopted is to generate a memory resident control flow graph (CFG) of input C program which represents semantically equivalent transformation of C program. Further, we traverse this CFG and compute abstract summary by generating invariants at different points in the program to check some properties of the original program. We analyze how abstract values for a set of program variables are updated by different constructs of C programs such as assignments, conditions, loops, etc. Our focus is on numerical properties of C programs. \\

\noindent The rest of the thesis is organized as follows:

\begin{itemize}
  \item Next chapter \ref{sec:LiteratureSurvey} Literature Survey describes some concepts and preliminaries to understand Abstract Interpretation.
  \item In chapter \ref{sec:CAnalyzer} C Analyzer Implementation, we describe plug-and-play domain architecture and implementation of C Analyzer tool built for static analysis of C programs.
  \item Chapter \ref{sec:ResultsDiscussions} Results and Discussions notes results of C Analyzer implementation and future work to be done in this direction.
  \item Chapter \ref{sec:SummaryConclusions} Summary and Conclusions summarizes this project work with concluding remarks.
\end{itemize}


\chapter{Literature Survey}
\label{sec:LiteratureSurvey}
In this chapter first, we will discuss some basic ideas and techniques related to lattice theory, abstract domains and Galois connection. These concepts are required to appreciate theory of Abstract Interpretation on which our tool C Analyzer is based.

\section{Lattice Theory}

\subsection{Partially Ordered Set}
A partially ordered set (poset) ($P, \le$) is a binary relation $\le$ over a set $P$ which is reflexive, antisymmetric and transitive. A partially ordered set formalizes the concept of ordering on the elements of a set. Partial order reflects the fact that not every pair of elements in the set need be related.


\subsection{Lattice}
A lattice is a partially ordered set (poset) in which any pair of elements has a supremum and an infimum. Supremum is also called a least upper bound (lub) or join. Infimum is also called a greatest lower bound (glb) or meet. \\

\noindent Formally, a poset ($L, \le $) is a lattice if it satisfies following axioms:
  \begin{itemize}
    \item For any two elements $a, b \in L$, the set \{a, b\} has a least upper bound or \textbf{lub} denoted as $a \vee b$.
    \item For any two elements $a, b \in L$, the set \{a, b\} has a greatest lower bound or \textbf{glb} denoted as $a \wedge b$.
  \end{itemize}

\noindent A lattice is said to \textbf{complete} if every subset $S$ of $L$ has an lub and a glb.


\subsection{Bounded Lattice}
A lattice is said to be bounded lattice if it has a greatest and a least element - also known as \textbf{Top} ($\top$) and \textbf{Bottom} ($\bot$) respectively. Any lattice ($L, \le $) can be converted to a bounded lattice ($ L, \le, \top, \bot $) by adding a greatest and a least element. The greatest element is obtained by taking join of all elements while the least element is obtained by taking meet of all elements. \\

\noindent The $\top$ and $\bot$ have following special properties:

\begin{itemize}
  \item $ \bot \wedge x = \bot $ and $ \bot \vee x = x    $, $ \; \forall x \in L $.
  \item $ \top \wedge x = x    $ and $ \top \vee x = \top $, $ \; \forall x \in L $.
\end{itemize}


\subsection{Ascending Chain Property}
A partially ordered set (poset) P is said to satisfy the ascending chain condition (ACC) if every strictly ascending sequence of elements eventually terminates, i.e. there is no infinite ascending chain \cite{WIKI}. \\

\noindent Formally, given any sequence
\begin{center}
$a_1 \le a_2 \le a_3 \le ...$, 
\end{center}
there exists a positive integer n such that
\begin{center}
$a_n = a_{n+1} = a_{n+2} = ...$. 
\end{center}


\subsection{Descending Chain Property}
A partially ordered set (poset) P is said to satisfy the descending chain condition (DCC) if every strictly descending sequence of elements eventually terminates, e.g. there is no infinite descending chain \cite{WIKI}. \\

\noindent Formally, given any sequence
\begin{center}
$... a_3 \le a_2 \le a_1$, 
\end{center}
there exists a positive integer n such that
\begin{center}
$... = a_{n+2} = a_{n+1} = a_n$. 
\end{center}

\noindent Note:- Both ascending chain and descending chain properties are finiteness properties for poset. Every finite poset satisfies both ACC and DCC \cite{WIKI}.


\section{Abstract Domains}

In this section, first we will discuss about program semantics and then abstract domains mentioned in this thesis. \\

\noindent Semantics of a program describes the set of all possible behaviors of the program when executed for all possible input data. A program behavior can be A) correct termination giving one or more output results, or B) termination in error condition, or C) non-termination. \\

\noindent For a given program, we talk about its concrete semantics and abstract semantics. \\

\noindent The concrete semantics of a program is an `infinite' mathematical object which is \emph{not computable}, i.e. it is not possible to write a program that is able to represent and compute all possible execution paths of any program in its all possible execution environments. Most of interesting program properties are \emph{undecidable} in concrete semantics. Hence concrete semantics of a program is mapped to a possible abstract semantics where program properties are decidable \cite{Intro-AI}. \\

\noindent Concrete semantics is described by a \emph{concrete domain} which is a set of all possible execution paths of a program in all possible execution environments. \\

\noindent Abstract semantics is described by an \emph{abstract domain} which is a semantic approximation covering all possible execution paths of a program. \\

\noindent We limit our discussion of an \emph{abstract domain} to computer recognizable program properties and a set of operators that manipulate them. Abstract domains considered in this thesis for numerical analysis, fall into following categories: Non-relational Numerical Abstract Domain and Relational Numerical Abstract Domain.


\subsection{Non-relational Numerical Abstract Domain}
Non-relational numerical abstract domain focuses on properties of individual numerical variables in a program, i.e. what values a program variable can take. \emph{Interval domain} is an example of non-relational numerical abstract domain.

\subsubsection{Interval}
Interval Abstract Domain is used to represent constraints of the form $  a \le x \le b $ where values of program variables are known to lie in certain interval only. This cannot represent relationship between values of two or more program variables.


\subsection{Relational Numerical Abstract Domain}
Relational Numerical Abstract Domain can discover relationship between program variables, i.e. are two variables a and b related by a constant c such that $\pm a \pm b \le c$. Examples of relational numerical abstract domain include \emph{Octagonal domain} and \emph{Polyhedron domain}. 

\subsubsection{Octagon}
Octagonal Abstract Domain is used to represent constraints of the form $\pm X \pm Y \le c$, where X and Y are program variables and c is a constant.


\subsubsection{Polyhedra}
Polyhedra Abstract Domain can infer linear relationships between variables, e.g. linear in-equalities of the form $ a_1 x_1 + a_2 x_2 + ...  + a_n x_n \le c $.


\subsection{Precision}

Figure \ref{fig:domains} below shows a high-level pictorial comparison of interval, octagonal and polyhedron domain. Figure has same set of points $\bullet$ abstracted in interval, octagonal and polyhedron domains and spurious points (false positives) are shown with $\times$ symbol.

\begin{figure}[h!]
  \centering
  \includegraphics[height=1in, width = 4.5in]{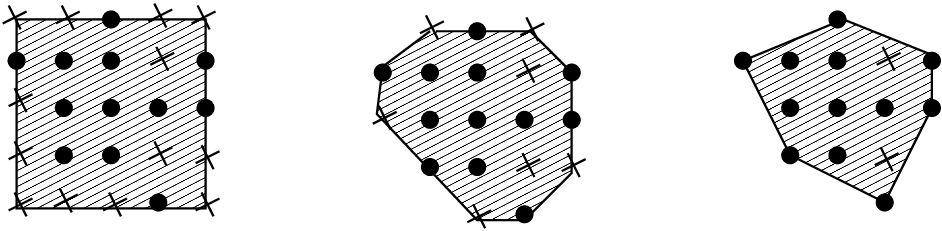}
  \caption{Comparison of Interval, Octagonal and Polyhedron Domains}
  \label{fig:domains}
\end{figure}

\noindent Interval analysis is very efficient with linear memory and time cost. It is faster but less precise than Octagonal domain which has $\mathcal{O}(n^2)$ worst case memory cost. Polyhedron analysis is much more precise but has a huge memory cost - exponential in number of variables. Precision and complexity of Octagonal lies between that of Interval and Polyhedron. These results are documented in \cite{LIT-Mine}.


\subsection{Tradeoff}
During analysis, we may have to choose precision over efficiency or efficiency over precision depending on analysis requirement. In some cases, we may end up using faster but imprecise Interval domain. On other occasions such as for floating points we may choose more precise but complex Polyhedra domain. A more precise domain comes with more memory and time cost associated. So, user can decide on some tradeoff between efficiency and precision to cater analysis needs, e.g. limit precision to gain on efficiency of analysis. \\

\noindent Interval domain gives range bound of individual numerical variables in a program. If we have to relate two program variables $ x $ and $ y $ by say $ x + y \le 10 $ then we cannot use Interval as it will only give range bound on these two variables, we need to use Octagon in this case. \\

\noindent Octagonal domain ($\pm X \pm Y \le c$) closely approximates to an octagon (polyhedra with at most eight sides) if lines are drawn in x-y plane. If some analysis requires us to analyze linear inequality of the form $ a_1 x_1 + a_2 x_2 + ...  + a_n x_n \le c $, we cannot use Octagon as that will be imprecise in this case and we need to use Polyhedra.


\section{Galois Connection}

\subsection{Galois Connection}
Galois connection is used to find a sound approximating abstract domain of a concrete domain and vis-a-versa. Galois connection is a particular correspondence between two partially ordered sets (poset). A partially ordered set is a binary relation over a set which is reflexive, antisymmetric and transitive. \\

\noindent Let ($P, \le$) and ($Q, \sqsubseteq$) be two poset. A pair ($\alpha, \gamma$) of maps $\alpha : P \rightarrow Q$ and $\gamma : Q \rightarrow P$ is called Galois connection iff $\forall x \in P, \forall y \in Q$,
\begin{equation}
\alpha(x) \sqsubseteq y \iff x \le \gamma(y)
\label{eq:gc1}
\end{equation}
written as
\begin{equation}
(P, \le) \stackrel{\overset{\alpha}{\longrightarrow}}{\underset{\gamma}{\longleftarrow}} (Q, \sqsubseteq)
\end{equation}

\begin{figure}[h!]
  \centering
  \includegraphics[scale=0.5]{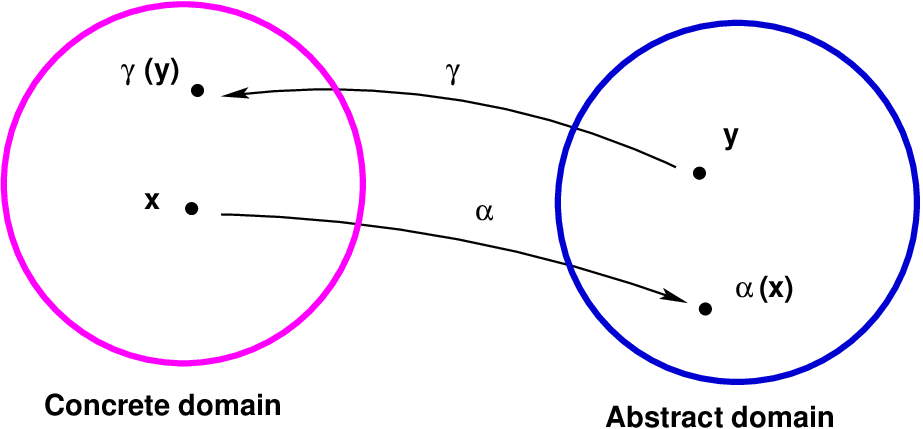}
  \caption{Galois connection}
  \label{fig:gc}
\end{figure}

\noindent Here $\alpha$ is \emph{Abstraction function} and $\gamma$ is \emph{Concretization function}. $\alpha$ is also called lower adjoint and $\gamma$ is called upper adjoint. Figure \ref{fig:gc} pictorially describes Galois connection equation \ref{eq:gc1}.

\noindent \emph{Why is partial order considered?} In general, it may not be possible to compare each and every element of domain or not all elements can be compared, (e.g. for integers minimum ($ -\infty $) and maximum ($ +\infty $) elements are not comparable), therefore without loss of generality, we consider partial order (and not total order). \\

\noindent \textbf{Example 1 - Interval Abstraction}
Let ($P, \le$) and ($Q, \le$) be two poset. \\
$P = \mathbb{Z} \cup \{+\infty\} \cup \{-\infty\}$ \\
$Q = \{[a, b] \mid a \in P, b \in P \}$ \\
$\alpha : P \rightarrow Q$ such that $\alpha(X) = [Min, Max], X \subseteq P$ \\
$\gamma : Q \rightarrow P$ such that $\gamma([Min, Max]) = \{ X \mid x \in X, Min \le x \le Max \}$ \\
Pair $(\alpha, \gamma$) forms a Galois connection. \\
e.g. \\
$X = \{2, 4, 6, 8, 10\}$ \\
$\alpha(X) = [2, 10]$ \\
$\gamma([2, 10]) = \{2, 3, 4, 5, 6, 7, 8, 9, 10\}$ \\

\noindent \textbf{Example 2 - Functional Abstraction}
Let ($P, \subseteq$) and ($Q, \subseteq$) be two poset. \\
$P = \mathbb{Z} $ \\
$Q = \{ -1, 0, +1 \}$ \\
$f(x) = (x < 0 \hspace{2pt}?\hspace{2pt} (-1) : (x = 0 \hspace{2pt}?\hspace{2pt} 0 : +1))$ \\
$\alpha : P \rightarrow Q$ such that $\alpha(X) = \{f(x) \mid x \in X \}$ \\
$\gamma : Q \rightarrow P$ such that $\gamma(Y) = \{x \mid f(x) \in Y \}$ \\
Pair $ (\alpha, \gamma$) forms a Galois connection from power set of P to power set of Q. \\
e.g. \\
$X = \{0, 1, 3, 5\}$ \\
$\alpha(X) = \{0, 1\}$ \\
$\gamma(\{0, 1\}) = \{ x \subseteq \mathbb{Z} \mid x \ge 0 \} = N$ i.e. all natural numbers.


\subsection{Properties of Galois Connection}
Let us see some interesting properties of Galois connection. \\

\noindent For simplicity and to understand properties, let ($P, \le$) and ($Q, \le$) be two poset. Pair ($\alpha, \gamma$) of maps $\alpha : P \rightarrow Q$ and $\gamma : Q \rightarrow P$ forms Galois connection iff $\forall x \in P, \forall y \in Q, \alpha(x) \le y \iff x \le \gamma(y)$ , then, following properties hold for Galois connection:

\begin{itemize}
  \item Compositions $\gamma(\alpha(x)) \ge x$ and $\alpha(\gamma(y)) \le y$: from defining property of Galois connection, we know that $\alpha(x) \le y \iff \gamma(y) \ge x$. Since $\alpha(x) \le \alpha(x)$, take $\alpha(x) = y$, then from right hand side of defining property: $\gamma(\hspace{5pt}\alpha(x)\hspace{5pt}) \ge x$ . Also, since $\gamma(y) \le \gamma(y)$, take $\gamma(y) = x$, then from left hand side of defining property: $\alpha(\hspace{5pt}\gamma(y)\hspace{5pt}) \le y$.

  \item $\alpha$ and $\gamma$ are monotonic: a function $f : P \rightarrow Q$ is monotonic, iff $\forall x, y \in P: (x \le y) \rightarrow f(x) \le f(y)$, i.e. order is preserved. So, we need to show if $x \le y$, then $\alpha(x) \le \alpha(y)$ and $\gamma(x) \le \gamma(y)$. Since $x \le y$ and from above property $\gamma(\alpha(y)) \ge y$, hence $x \le y \le \gamma(\hspace{5pt}\alpha(y)\hspace{5pt})$. From this and left hand side of defining property, we get $\alpha(\hspace{5pt}x\hspace{5pt}) \le \alpha(y)$. Since $x \le y$ and from above property $\alpha(\gamma(x)) \le x$, hence $\alpha(\hspace{5pt}\gamma(x)\hspace{5pt}) \le x \le y$. From this and right hand side of defining property, we get $\gamma(x) \le \gamma(\hspace{5pt}y\hspace{5pt})$.

  \item $\alpha(\gamma(\alpha(x))) = \alpha(x)$ and $\gamma(\alpha(\gamma(x))) = \gamma(x)$: from first property $\gamma(\alpha(x)) \ge x$ and $\alpha(\hspace{5pt}\gamma(\alpha(x))\hspace{5pt}) \ge \alpha(\hspace{5pt}x\hspace{5pt})$ (taking $\alpha$ both sides since $\alpha$ is monotonic). Also from first property, $\alpha(\gamma(y)) \le y$, take $\alpha(x) = y$ gives $\alpha(\gamma(\hspace{5pt}\alpha(x)\hspace{5pt})) \le \alpha(x)$. Hence it follows $\alpha(\gamma(\alpha(x))) = \alpha(x)$. Similarly we can show $\gamma(\alpha(\gamma(x))) = \gamma(x)$.

\end{itemize}


\subsection{Galois Insertion}
A Galois connection is called a Galois Insertion if:
\begin{itemize}
  \item $\gamma(\alpha(x)) = x$ , identity function or
  \item $\gamma$ is one-to-one or
  \item $\alpha$ is onto
\end{itemize}

\noindent \emph{Why Galois Insertion is important?} It may not be possible to find Galois Insertion for a Galois connection always. But when we can find Galois Insertion, it minimize false positives and gives best approximating solution.


\section{Abstract Interpretation}
\label{sec:AbstractInterpretation}

The core idea of Abstract Interpretation is formalization of notion of approximation. Initially approximation of memory configurations (e.g. program variables) is defined. Then approximation of all atomic operations (e.g. arithmetic, relational operations) is defined. Further approximation is lifted to entire program structure \cite{ABS-INT-TUT} (e.g. using abstract operators).  \\

\noindent We start with a formal specification of the program semantics (program variables in \emph{Concrete semantic}). Then we construct abstract semantic equations with respect to a parametric approximation scheme. Use general algorithms to solve these abstract semantic equations. Then we try to find best-fit approximation that suits the purpose \cite{ABS-INT-TUT}.


\subsection{Collecting Semantics}
Collecting semantics is the set of observable behaviors (or all the states) defined by operational semantics of structure of the program. It is the starting point of the analysis. It means finding initial state (entry point of the program), set of all descendent states of the initial state (all program points reachable from entry point), set of all finite paths that can reach a final state (exit point of the program), etc \cite{ABS-INT-TUT}. \\

\noindent \emph{What are we collecting}:

  \begin{itemize}
    \item state properties: divide by zero, overflow, etc.
    \item finite and infinite path properties: uninitialized variable being used, termination of loop, etc.
  \end{itemize}

\noindent In Abstract Interpretation the collecting semantics of a program is expressed as a least fix-point of a set of equations. The equations are solved over some abstract domain that captures the property to be analyzed. The equations are solved iteratively i.e. successive approximation of the solution is computed until a fix-point is reached \cite{WIDNARO}.


\section{Abstract Operators}
Common abstract operators are: meet, join, widening, narrowing of two given abstract values. These operators are binary operators.

\subsection{Meet}
Meet of two abstract values is greatest lower bound (glb) in lattice. \\

\noindent Formally, for a lattice ($L, \le$), an element $ z \in L $ is meet of two elements $ x $ and $ y $, if 

\begin{itemize}
  \item z is lower bound of $ x $ and $ y $: $ z \le x$ and $ z \le y $. 
  \item z is greater than or equal to any other lower bound on $ x $ and $ y $: for any $ z \in L $, such that $ w \le x $ and $ w \le y $, then $ w \le z $.
\end{itemize}

\noindent Also, $ \bot \wedge x = \bot $ and $ \top \wedge x = x $, $ \; \forall x \in L $.


\subsection{Join}
Join of two abstract values is least upper bound (lub) in lattice. \\

\noindent Formally, for a lattice ($L, \le$), an element $ z \in L $ is join of two elements $ x $ and $ y $, if

\begin{itemize}
  \item z is upper bound of $ x $ and $ y $: $ z \ge x$ and $ z \ge y $. 
  \item z is less than or equal to any other upper bound on $ x $ and $ y $: for any $ z \in L $, such that $ w \ge x $ and $ w \ge y $, then $ w \ge z $.
\end{itemize}

\noindent Also, $ \bot \vee x = x $ and $ \top \vee x = \top $, $ \; \forall x \in L $. \\

\noindent Note:- From above discussions, we find that a) join and meet are dual binary operations and b) join has tendency to take us higher in the lattice while meet takes us lower in the lattice.


\subsection{Widening}
In Abstract Interpretation, approximation of the solution is computed iteratively until a fix-point is reached. For some abstract domains, such chains can be either infinite or too long to have the analysis efficient. To work with these domains, Abstract Interpretation provides a powerful tool - widening operators that attempt to predict the fix-point based on the sequence of approximations computed on earlier iterations of the analysis on a complete lattice \cite{WIDNARO}. \\

\noindent Formally, for lattice ($L, \le$): a widening operator $ \bigtriangledown $ is a function $ \bigtriangledown : L \times L \rightarrow L $ such that
\begin{center} $ \forall x, y \in L : x \le (x \bigtriangledown y) \; \; \& \; \; y \le (x \bigtriangledown y) $ \end{center} 
\noindent and it stabilizes after a fixed number of terms for $ n \ge 0 $, for ascending chain defined as
$$
\left \{
\begin{array}{lr}
Y_0 = X_0 \\
Y_{n+1} = Y_n \bigtriangledown X_{n+1}
\end{array}
\right.
$$

\noindent Hence $ (Y_n)_{n \ge 0} $ eventually converges to fix point. \\

\noindent Note:- Number of iterations required to reach fix point may depend on the library or user implementing widening. \\

\noindent Widening for Intervals $ [a_0, b_0]  \bigtriangledown [a_1, b_1] $ is defined as below:
\begin{center}
If $ a_0 \le a_1 $ then $ a_0 $ else $ -\infty $ \\
If $ b_1 \le b_0 $ then $ b_0 $ else $ +\infty $  \\
$ x \bigtriangledown \bot = \bot \bigtriangledown x = x $ \\
$ x \bigtriangledown \top = \top \bigtriangledown x = \top $
\end{center}

\noindent \textbf{Examples - Widening}:
\begin{center}
$ [2, 3] \bigtriangledown [1, 4] = [-\infty, +\infty] $ \\
$ [0, 1] \bigtriangledown [0, 2] = [0, +\infty] $ \\
$ [1, 4] \bigtriangledown [2, 3] = [1, 4] $
\end{center}

\noindent Widening may degrade precision of the solution due to faster convergence to fix-point. This can be offset by some optimizations like unrolling a loop by $n$ times to delay widening for successive iterative approximation or using narrowing operators.


\subsection{Narrowing}
The degradation of precision of the solution obtained by widening operator can be partly restored by further applying a narrowing operator. Narrowing operators soundly improves precision of an approximation obtained with widening operator. Widening may have introduced infinite bounds for faster convergence to fix point. Narrowing operator improves infinite bounds whenever possible \cite{AI-SUTRE}. \\

\noindent Formally, let ($L, \le$) be a lattice. A narrowing operator $\bigtriangleup$ is a function $ \bigtriangleup : L \times L \rightarrow L $ such that
\begin{center} $ \forall x, y \in L : x \le y \Longrightarrow ( x \le (y \bigtriangleup x) \le y ) $ \end{center} 
\noindent and it stabilizes after a fixed number of terms for $ n \ge 0 $, for decreasing chain defined as
$$
\left \{
\begin{array}{lr}
Y_0 = X_0 \\
Y_{n+1} = Y_n \bigtriangleup X_{n+1}
\end{array}
\right.
$$

\noindent Narrowing for Intervals $ [a_0, b_0] \bigtriangleup [a_1, b_1] $ is defined as below:
\begin{center}
If $ a_0 = -\infty $ then $ a_1 $ else $ a_0 $ \\
If $ b_0 = +\infty $ then $ b_1 $ else $ b_0 $ \\
$ x \bigtriangleup \bot = \bot \bigtriangleup x = \bot $ \\
$ x \bigtriangleup \top = \top \bigtriangleup x = x $
\end{center}

\noindent \textbf{Examples - Narrowing}:
\begin{center}
$ [-\infty, +\infty] \bigtriangleup [-\infty, 101] = [-\infty, 101] $ \\
$ [1, +\infty] \bigtriangleup [50, 100]  = [1, 100] $ \\
$ [1,4] \bigtriangleup [2, 3] = [1, 4] $
\end{center}

\vspace{1cm}
\noindent Next chapter \ref{sec:CAnalyzer} C Analyzer Implementation describes C Analyzer tool and contributions of this project work.


\chapter{C Analyzer Implementation}
\label{sec:CAnalyzer}

In this chapter, we describe various ideas related to implementation of C Analyzer and major contributions of this thesis. C Analyzer leverages basic ideas and techniques covered under previous chapter \ref{sec:LiteratureSurvey} Literature Survey. This tool is based on the theory of Abstract Interpretation to compute approximate analysis of the program. First we provide a high level overview of the tool, followed by CIL transformations and data structures used. Further we describe general processes and algorithms used during analysis and features supported by C Analyzer. Later, we show how to add a new domain for analysis.

\section{Overview}

\begin{figure}[h!]
  \centering
  \includegraphics[scale=1]{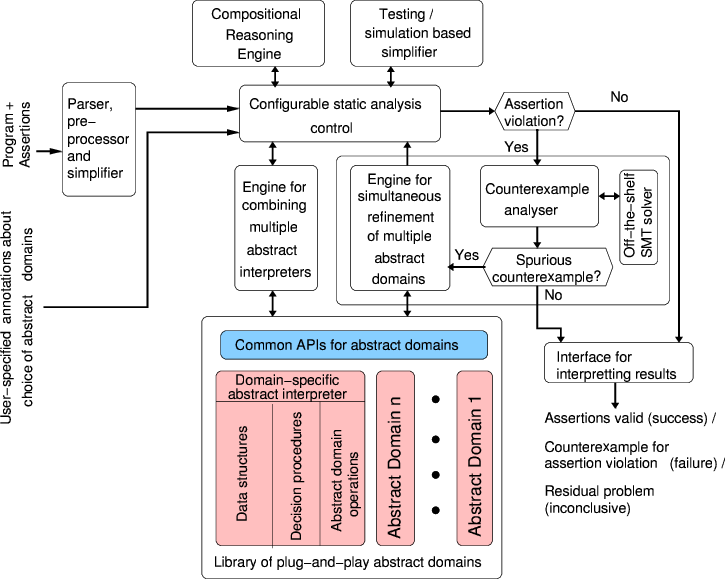}
  \caption{CFDVS Project Proposal - Block Diagram}
  \label{fig:proposal}
\end{figure}

\noindent As per the referenced block diagram \ref{fig:proposal} taken from CFDVS project proposal report, we need the support for multiple abstract domain to be used in plug-and-play manner so that in future multiple abstract domains can be combined to provide more precise analysis and abstraction refinement techniques can be used with these domains. \\

\noindent C Analyzer is a tool built for static analysis of C programs using theory of Abstract Interpretation. C Analyzer provides a plug-and-play architecture for multiple abstract domains to be used for this requirement and we are not dependent on a particular library provided for any abstract domain. We have a common interface for analysis to be used across domains and we use virtual polymorphism to invoke domain specific implementation depending on domain selected by user. \\

\noindent We use CIL (Common Intermediate Language), a source-to-source transformation tool, to transform some features in C programs to simpler constructs, such as Logical operators, switch statement, conditional operator etc. are transformed into if-else ladder. \\

\noindent C Analyzer uses LLVM's C/C++ compiler Clang v3.1 for CFG generation and traversal. Clang provides rich API to read AST while iterating over statements in basic blocks of CFG during CFG traversal. Using C Analyzer we generate invariants for program statements in different abstract domains. These invariants help to verify implicit assertions (e.g. divide by zero) and user defined explicit assertions ($ x \textgreater 0 $) for analyzing certain properties of the program. \\

\noindent Currently following abstract domains are supported by C Analyzer:
  \begin{itemize}
    \item Box (or interval) - invariants are represented in the form of interval [min, max] for values for variables in the program.
    \item Octagon - invariants are in the form of equations for related variables in the program.
    \item Polyhedra - invariants are in the form of linear equations for variables in the program.
    \item Bit Vector - invariants are represented by symbolic expressions (work to use this domain and to be plugged fully in plug-and-play architecture is under progress).
  \end{itemize}

\noindent Abstract domains Box (or Interval), Octagon, Polyhedra are provided by Apron v0.9.9 \cite{APRON} and Bit Vector domain library is provided by another project in CFDVS. \\

\noindent Note:- LLVM and its subproject Clang are released under University of Illinois/NCSA Open Source License. APRON is released as a free software under LGPL license. CIL is released under BSD open source license.


\section{CIL Transformations}
We use CIL (Common Intermediate Language) as a source-to-source transformation tool. Using CIL following C constructs are simplified into simpler C construct if-else ladder:

  \begin{itemize}
    \item Logical AND and OR operators
    \item Switch statement
    \item Conditional operator
  \end{itemize}

\noindent Using CIL, do-while and for loops are transformed into while loops that is supported by C Analyzer. Break statements are transformed into a goto statement and a labelled null statement. \\

\noindent By doing these source-to-source transformations we reduce number of C constructs we need to address (writing visitor methods to read AST for statements involving these operators or statements) using Clang API. \\

\noindent CIL removes variables declared but unused anywhere in the program. CIL also simplifies some numerical expressions involving constants, e.g. x = 1+1; is reduces to x = 2; \\

\noindent One side effect of using CIL is, it introduces C style explicit casts. Therefore, support for C style explicit cast expressions has been added \ref{sec:CStyleExplicitCasts}.


\section{C Analyzer: Data Structures and Classes}
\label{sec:CAnalyzerClasses}

This section describes data structures and classes created for C Analyzers and some important functions in the execution flow of this tool.

\subsection{Data Structures}

\subsubsection{WrapperAbsVal}
WrapperAbsVal is structure to wrap abstract values for a domain. This contains two members: 

\begin{itemize}
  \item Aval: a generic pointer (void *) to wrap domain specific abstract value to be passed on to or receive abstract values from a common interface across domains
  \item domain: an integer identifier for abstract domains; 1 - Interval, 2 - Octagon, 3 - Polyhedra and 4 - Bit Vector. This can be used for sanity check of abstract values being passed to and received from common interface to ensure abstract value belongs to intended domain.
\end{itemize}


\subsubsection{MyCFGInfo}
\label{sec:MyCFGInfo}
MyCFGInfo is structure to store abstract summary at the end of a basic block for all CFG blocks. MyCFGInfo contains block ID, pointer to basic block, terminator type of block (Empty (entry and exit blocks), None (block with no terminator), If (source of condition), While (source of loop), etc.), a flag to denote if this block is source of a back edge, abstract value at the end of the block, abstract value of positive and negative of condition of the block, previous abstract value at loop exit. Any abstract value is empty if not applicable for a basic block in MyCFGInfo.

\begin{figure}[ht]
\centering
  \includegraphics[scale=0.6, angle=0]{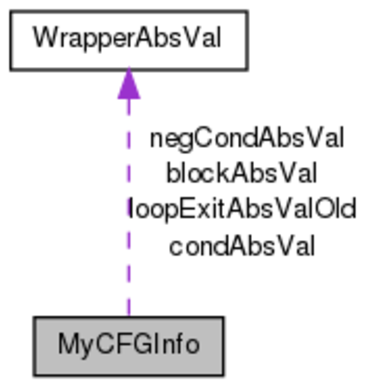}
\caption{MyCFGInfo: Collaboration Diagram}
\label{fig:MyCFGInfoCol}
\end{figure}

\noindent Note:- Abstract values stored in MyCFGInfo are wrapped inside a structure called WrapperAbsVal described above.


\subsubsection{MyCFGInfoList}
\label{sec:MyCFGInfoList}
MyCFGInfoList is a structure to store list of MyCFGInfo entries for all basic blocks of CFG during analysis. Essentially, MyCFGInfoList contains a vector of pointers to MyCFGInfo structures. 


\subsection{Driver Program}

CFGGenerator is the driver program to get input C program and a file name to dump analysis details. CFGGenerator sets compiler instance options. Compiler instance is an instance of Clang class CompilerInstance that manages various objects - preprocessor, platform specific target information, language options, and ASTContext, etc. and provides utility function to manage clang objects. An instance of compiler instance must be active all time during Clang execution flow. \\

\noindent CFGGenerator also gets HEADER\_SEARCH\_PATH for input C program headers. This is a environment variable to be set before running executable CAnalyzer and it contains colon separated paths for headers. CFGGenerator creates file manager, source manager, preprocessor, ASTContext (keeps AST node types and declarations), AST reader (MyASTConsumer) and invokes parser by calling ParseAST(). See llvm-3.1.src/tools/clang/lib/Parse/ParseAST.cpp file for ParseAST() and flow from there on.


\subsection{Classes}

\subsubsection{MyASTConsumer}

\begin{figure}[ht]
\centering
  \includegraphics[scale=0.6, angle=270]{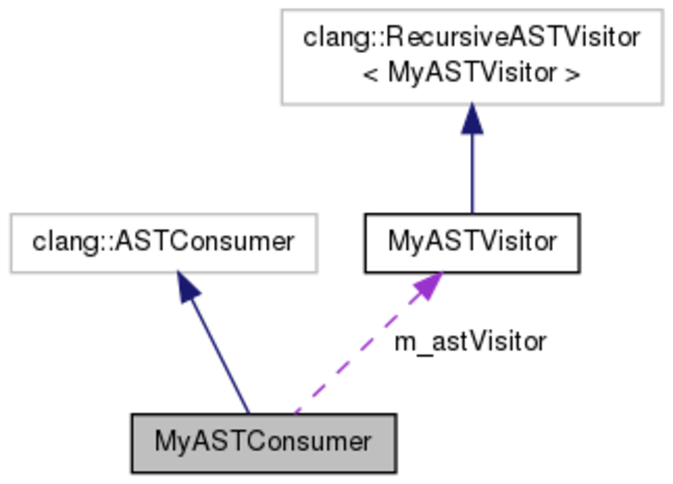}
\caption{MyASTConsumer: Collaboration Diagram}
\label{fig:MyASTConsumerCol}
\end{figure}

\noindent MyASTConsumer is AST reader class, inherited from Clang class ASTConsumer. MyASTConsumer is instantiated by CFGGenerator. This sets compiler instance for self, creates instance of AST visitor - MyASTVisitor and sets compiler instance for MyASTVisitor. When CFGGenerator calls ParseAST(), this will call HandleTopLevelDecl() on AST reader (MyASTConsumer) object. HandleTopLevelDecl() is a virtual function which is overridden by MyASTConsumer to call TraverseDecl() on MyASTVisitor instance in order to visit every top declaration in the input C program (function declarations, function definitions, global variables, structure and such other global declarations).







\subsubsection{MyASTVisitor}

\begin{figure}[ht]
\centering
\includegraphics[scale=0.6, angle=270]{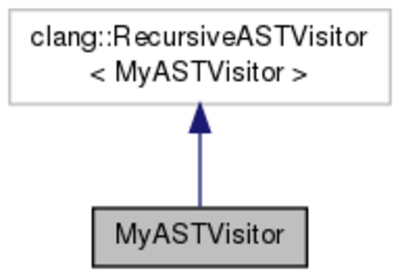}
\caption{MyASTVisitor: Inheritance Diagram}
\label{fig:MyASTConsumerInherit}
\end{figure}

\noindent MyASTVisitor is AST node visitor class, inherited from Clang's template class RecursiveASTVisitor. RecursiveASTVisitor is a very important template class to be used to leverage Clang API and as name suggests this recursively visits each AST node from top to bottom. \\

\noindent Any class inheriting from RecursiveASTVisitor can override visitor functions of interests, e.g. MyASTVisitor overrides VisitFunctionDecl(). When MyASTConsumer calls TraverseDecl() on MyASTVisitor instance, this in turn calls VisitFunctionDecl(). By overriding VisitFunctionDecl(), MyASTVisitor instantiates MyCFG class, creates CFG for a function definition by calling getCFG() on instance of MyCFG. \\

\noindent After CFG is created, MyASTVisitor is responsible for pre-processing of CFG before analysis begins on list of CFG blocks. There are two primary tasks for pre-processing:

\begin{itemize}
  \item create blockList - a list of basic blocks of CFG to be visited in order
  \item create edgeMatrix - create a 2D vector (a map for source and destination blocks) containing edge information for basic blocks of CFG
\end{itemize}

\noindent Note:- Current logic to create blockList (taking care of back edges for loops and with changes for break and goto statements while using CIL), was implemented by Prateek Saxena (BARC). \\

\noindent After pre-processing of CFG, MyASTVisitor asks for choice of domain from user. Depending on domain selected, MyASTVisitor instantiates domain class (inherited from Analyzer) which calls constructor of parent class Analyzer to pass compiler instance, function declaration object, CFG object, blockList and edgeMatrix to Analyzer. Afterwords domain class does initialization of its internal data structures. \\

\noindent Further, it makes pointer of base class Analyzer, point to object of derived domain class object. This is used to take advantage of virtual polymorphism - by pointing base class object to derived class object, at run time implementation (for virtual functions of base class) in derived domain class is invoked based on which domain object is being pointed to by Analyzer. Finally, MyASTVisitor calls processCFG() on Analyzer to start CFG traversal for analysis.


\subsubsection{MyCFG}
MyCFG is wrapper class for memory resident CFG object being created. This class gets compiler instance and has a method getCFG() called by MyASTVisitor.VisitFunctionDecl(). Function getCFG() calls clang::CFG::buildCFG() which returns a pointer to CFG created. MyCFG gets this memory resident CFG object to be used by MyASTVisitor as described above.


\subsubsection{Analyzer}

\begin{figure}[ht]
\centering
  \includegraphics[scale=0.6, angle=270]{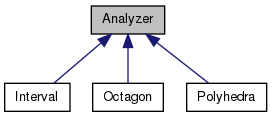}
\caption{Analyzer: Inheritance Diagram}
\label{fig:AnalyzerInherit}
\end{figure}

\noindent Analyzer class acts as a common interface across different abstract domains. It uses structure WrapperAbsVal containing generic pointer (void *) to wrap domain specific abstract values to be passed on to or receive abstract values from overridden virtual functions of base class Analyzer in the domain specific implementation of these functions. \\

\noindent Domain specific classes inherit from Analyzer as depicted in diagram \ref{fig:AnalyzerInherit}. \\

\noindent There are three types of virtual functions in Analyzer:

\begin{itemize}
  \item MyCFGInfo related: functions to maintain MyCFGInfo and get details from MyCFGInfo structure.
  \item Analyzer related: these virtual functions will be overridden for domain specific implementation to be called from inside processCFG() of Analyzer during CFG traversal.
  \item MyProcessStmt related: these functions are called from within visit methods of MyProcessStmt class while iterating over statements inside a basic block.
\end{itemize}


\subsubsection{MyProcessStmt}

\begin{figure}[ht]
\centering
  \includegraphics[scale=0.6, angle=270]{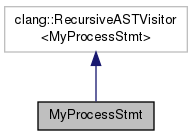}
\caption{MyProcessStmt: Inheritance Diagram}
\label{fig:MyProcessStmtInherit}
\end{figure}

\noindent MyProcessStmt is statement level processing class, inherited from RecursiveASTVisitor. MyProcessStmt object is called for every C statement inside a basic block during CFG traversal with pointer to Analyzer. This pointer to Analyzer is used to resolve domain type being pointed to at run time for appropriate virtual function implementations to be called from inside MyProcessStmt. \\

\noindent As mentioned earlier any class inheriting from RecursiveASTVisitor can override visitor functions of interests. MyProcessStmt overrides several C expressions and statements specific visit functions to get details from AST for each expression or statement including but not limited to all declarations, binary and unary expressions, conditions, loop statements, function calls, etc. \\

\noindent Phrase visit functions or visitors comes from the fact that these functions are names as Visit\#\#, where \#\# is replaced by statement class corresponding to declarations, expressions, conditional statements, loop statements, etc. e.g. VisitDeclStmt(), VisitBinAssign, VisitIfStmt(), etc. See \cite{StmtRef} for more such statement classes and \cite{RecASTV} for more visitor functions.


\subsubsection{Interval}

\begin{figure}[ht]
\centering
  \includegraphics[scale=0.6, angle=0]{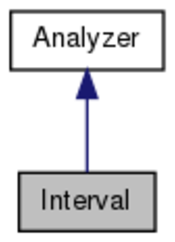}
\caption{Interval: Inheritance Diagram}
\label{fig:IntervalInherit}
\end{figure}

\noindent Interval is box (interval) domain specific implementation class. This is inherited from Analyzer class and defines interval's implementation for virtual functions of Analyzer along with its own internal data structures (manager, environment, expression stack, etc.) and functions.


\subsubsection{Octagon}

\begin{figure}[ht]
\centering
  \includegraphics[scale=0.6, angle=0]{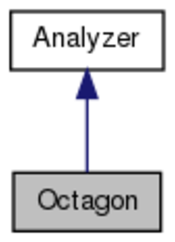}
\caption{Octagon: Inheritance Diagram}
\label{fig:OctagonInherit}
\end{figure}

\noindent Octagon is octagonal domain specific implementation class. This is inherited from Analyzer class and defines octagon's implementation for virtual functions of Analyzer along with its own internal data structures (manager, environment, expression stack, etc.) and functions.


\subsubsection{Polyhedra}

\begin{figure}[ht]
\centering
  \includegraphics[scale=0.6, angle=0]{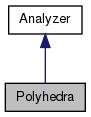}
\caption{Polyhedra: Inheritance Diagram}
\label{fig:PolyhedraInherit}
\end{figure}

\noindent Polyhedra is polyhedra domain specific implementation class. This is inherited from Analyzer class and defines polyhedra's implementation for virtual functions of Analyzer along with its own internal data structures (manager, environment, expression stack, etc.) and functions. \\

\noindent Note:- Abstract domain's internal implementation for Interval, Octagon, Polyhedra is provided by Apron library.


\subsubsection{BitVector}
BitVector is bit vector domain specific implementation class. This is inherited from Analyzer class and defines bit vector's implementation for virtual functions of Analyzer along with its own internal data structures (manager, environment, expression stack, etc.) and functions. \\

\noindent Note:- Bit Vector's internal implementation is provided by another project in CFDVS. This is being used by Prateek Saxena (BARC) for abstract analysis and work on usage of this domain and to fully plug into current plug-and-play architecture is under progress.


\pagebreak

\section{General Processes and Algorithms During Analysis}

This section describes some of the processes and algorithms used during analysis.

\subsection{CFG Generation}

\begin{figure}[htb]
\centering
\includegraphics[scale=0.5]{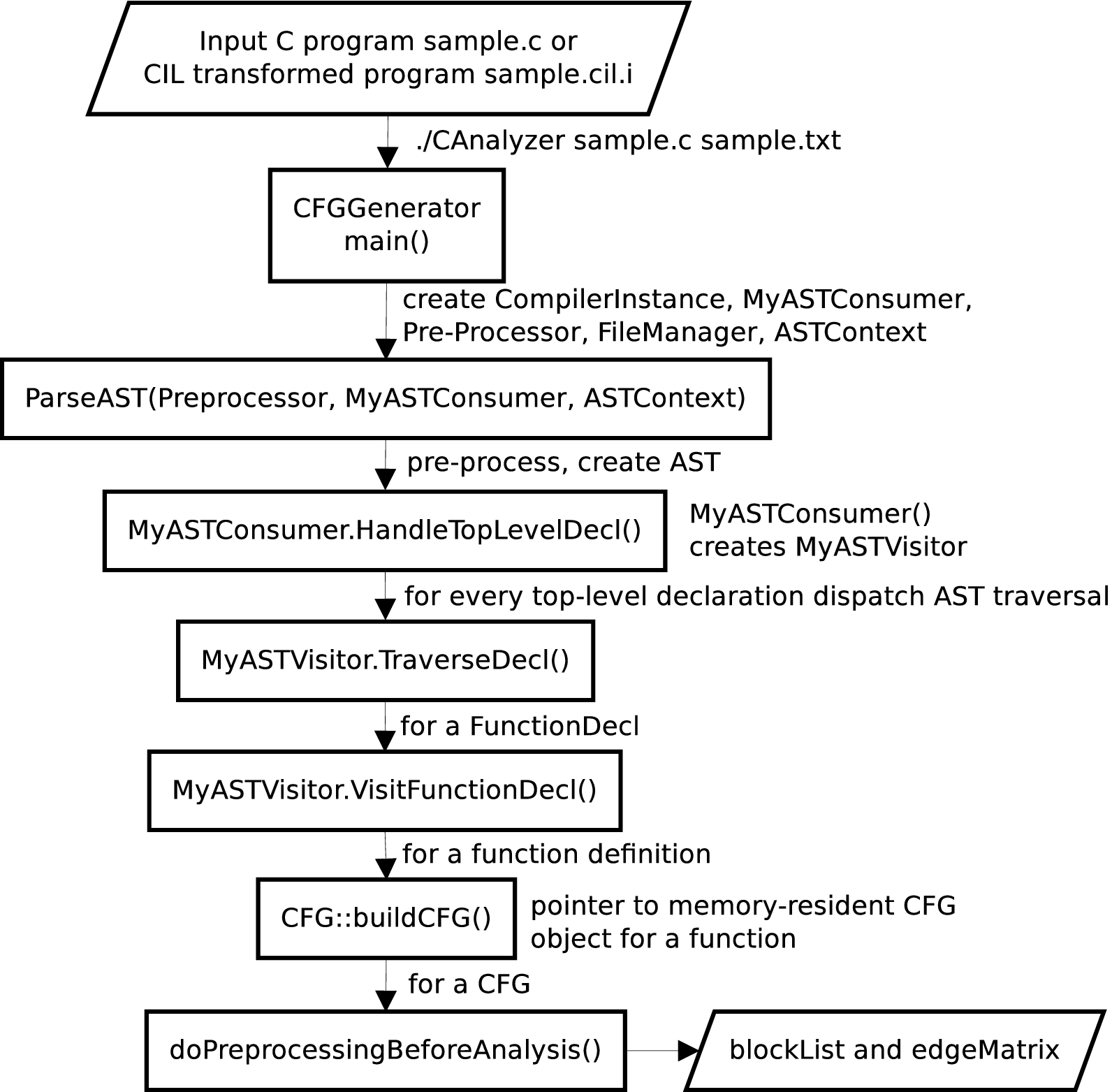}
\caption{CFG Generation}
\label{fig:cfg_gen}
\end{figure}

Figure \ref{fig:cfg_gen} a box and line diagram summarizes CFG generation in C Analyzer's execution flow. \\

\noindent CFGGenerator's main method in C Analyzer, takes a C program and a file name (to log analysis details) as input. CFGGenerator creates CompilerInstance, MyASTConsumer (AST reader), etc. and invokes parser by calling ParseAST() with MyASTConsumer instance as argument. \\

\noindent MyASTConsumer creates MyASTVisitor instance and overrides HandleTopLevelDeclaration() to dispatch AST traversal for every top level declaration (e.g. function definition) which in turn calls VisitFunctionDeclaration(). This creates CFG and does pre-processing before actual analysis, creates blockList - a list of blocks to be visited in order and edgeMatrix - a 2D vector to store edge information. \\

\noindent Note:- C Analyzer classes are described in detail under section \ref{sec:CAnalyzerClasses}.


\subsection{CFG Iteration}

\begin{figure}[htb]
\centering
\includegraphics[scale=0.5]{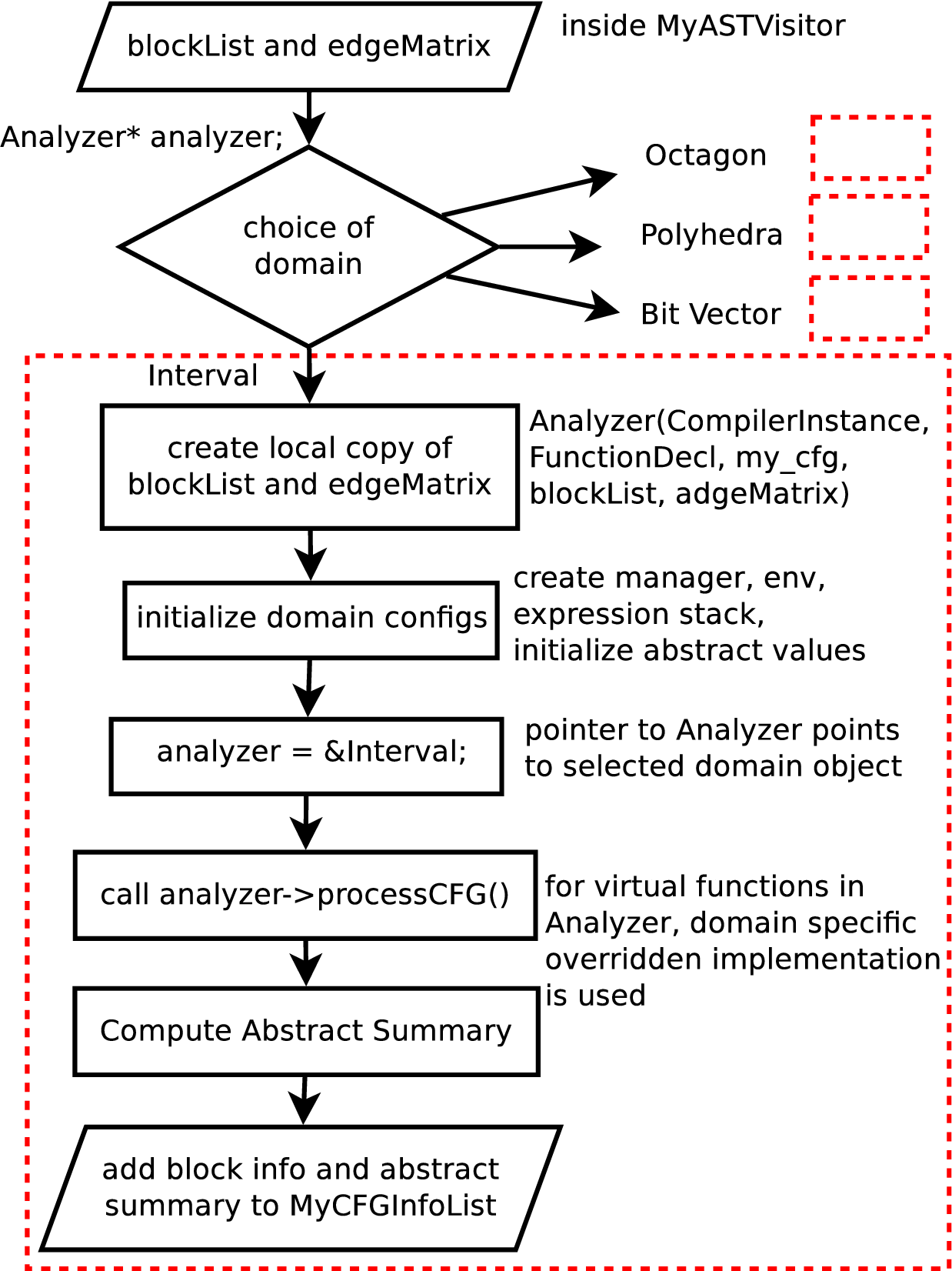}
\caption{CFG Iteration}
\label{fig:cfg_proc}
\end{figure}

Figure \ref{fig:cfg_proc} describes CFG traversal in C Analyzer's execution flow. \\

\noindent In MyASTVisitor, output of CFG generation blockList and edgeMatrix become input for CFG iteration. For a domain selected by user, domain class is instantiated which first calls parent class constructor Analyzer(CompilerInstance, FunctionDecl, my\_cfg, blockList, edgeMatrix) to create local copy of blockList and edgeMatrix and then initializes domain configurations - creating manager, environment, initialize abstract values, etc. \\

\noindent Now, pointer to Analyzer class points to derived domain class to leverage virtual polymorphism. Then Analyzer calls processCFG() to start CFG traversal for all basic blocks of CFG and computes abstract summary at the end of each block which is stored into MyCFGInfoList. \\

\noindent Note:- this flow is also described while C Analyzer classes are introduced under section \ref{sec:CAnalyzerClasses}.


\subsection{Computing Abstract Summary}
To compute abstract summary after every basic block in CFG, we use following algorithm: 

\begin{algorithm}
\label{sec:AbstractSummary}
\caption{Algorithm to compute abstract summary during analysis}
\begin{algorithmic}[1]

\Procedure{Generate-Traversal-Order}{{CFG} $\mathcal{G}$}
\State Return $L_{\mathcal{G}}$ - list of basic blocks to be visited in order for CFG $\mathcal{G}$.
\EndProcedure
\Statex


\Procedure{ProcessCFG}{{CFG} $\mathcal{G}$}
\For {Each block $\B$ in $L_{\mathcal{G}}$}

\State Let $PredList(\B)$ be the predecessor list of $\B$.

\If {($|PredList(\B)| > 1$)}
\State Abstract value at the beginning of $\B$ = $\mathsf{JoinBefore}(\B)$

\ElsIf {($|PredList(\B)| = 1$)}

\If {terminator type is If or While for block $\B'$ in $PredList(\B)$}
\State Abstract value at the beginning of $\B$ = $\mathsf{MeetBefore}(\B, \B')$

\ElsIf {$B$ is unique successor $\B'$ in $PredList(\B)$}
\State Abstract value at the beginning of $\B$ = block abstract value of predecessor block $\B'$

\EndIf

\EndIf

\State $\mathsf{ProcessStmt}(\B)$

\If {(Current block $B$ is a source of back edge)}
\State $\mathsf{WidenAbsVal}(\B)$
\EndIf

\State Add abstract value(s) at the end of the block into MyCFGInfoList[\ref{sec:MyCFGInfoList}]

\EndFor
\EndProcedure
\Statex


\Procedure{JoinBefore}{{CFG Block} $\B$}
\For {Each block $\B'$ in $PredList(\B)$ in order}
\State edge = edge from predecessor $\B'$ to $\B$
\If {((edge is not back edge or edge has been visited)} 
\State JoinedAbsVal = block abstract value of $\B'$
\EndIf
\State Mark edge as visited
\State index = index of block $\B'$ in $PredList(\B)$
\EndFor

\For {Each block $\B'$ in $PredList(\B)$ from index + 1 to end of $PredList(\B)$}
\State edge = edge from predecessor $\B'$ to $\B$
\If {((edge is not back edge or edge has been visited)} 
\State JoinValFromPred = block abstract value of $\B'$
\State JoinedAbsVal = JoinedAbsVal $\vee$ JoinValFromPred
\EndIf
\State Mark edge as visited
\EndFor

\State Return JoinedAbsVal.
\EndProcedure
\Statex


\Procedure{MeetBefore}{{CFG Block} $\B$, {CFG Block} $\B'$}

\If {$\B$ is first successor of predecessor $\B'$}
\State Abstract value at the beginning of $\B$ = block abstract value of $\B'$ $ \wedge $ positive of condition's abstract value of $\B'$
\Else
\State Abstract value at the beginning of $\B$ = block abstract value of $\B'$ $ \wedge $ negative of condition's abstract value of $\B'$

\algstore{algo1}
\end{algorithmic}
\end{algorithm}


\begin{algorithm}
\ContinuedFloat
\caption{Algorithm to compute abstract summary during analysis (continued...)}
\begin{algorithmic}
\algrestore{algo1}

\EndIf
\EndProcedure
\Statex


\Procedure{ProcessStmt}{{CFG block} $B$}

\For {Each statement $s$ in block $\B$}
\If {is $s$ last statement of block $\B$}
\State Skip (do nothing)
\Else
\State $\mathsf{MyProcessStmt.TraverseStmt}(s)$ and update abstract summary of the block $\B$
\EndIf
\EndFor

\State get terminator $t$ of block $\B$
\State $\mathsf{MyProcessStmt.TraverseStmt}(t)$ and update abstract summary of the block $\B$

\EndProcedure
\Statex


\Procedure{WidenAbsVal}{{CFG block} $B$}

\If{(is this first time visit to block $B$)}
\State previous abstract value at loop exit = $ \bot $
\Else
\State get previous abstract value at loop exit for block $B$ from MyCFGInfoList[\ref{sec:MyCFGInfoList}]
\EndIf

\State set current abstract value at loop exit = abstract value at the end of block $B$

\If {Back edge has been visited more than once} 

\If {number of times back edge visited = NUM\_UNROLLINGS}
\State new abstract value at loop exit = previous abstract value at loop exit $ \bigtriangledown $ current abstract value at loop exit
\Else
\State new abstract value at loop exit = current abstract value at loop exit
\EndIf

\If {new abstract value at loop exit = previous abstract value at loop exit}
\State fix-point reached
\State reset number of times back edge visited for block $B$
\Else
\State previous abstract value at loop exit = new abstract value at loop exit
\EndIf

\Else 

\State previous abstract value at loop exit = current abstract value at loop exit
\State new abstract value at loop exit = current abstract value at loop exit
\EndIf

\State set current abstract value at loop exit  = new abstract value at loop exit

\EndProcedure

\end{algorithmic}
\end{algorithm}



\pagebreak

\section{Features Implemented}

Apart from providing plug-and-play multiple domain architecture, this section describes our major contributions for features supported so far. Current implementation of C Analyzer supports following features of C constructs:

\subsection{Declarations}
All types of declarations related to characters, integers (both signed and unsigned), real numbers and declarations with initial values assigned are taken care of by VisitDeclStmt() inside MyProcessStmt class. Each abstract domain is supposed to add variables for these declaration to a vector or environment to keep track of them while iterating over basic blocks. \\

\noindent For Interval, Octagon, Polyhedra domains following keywords are supported for declarations: int, const (int), signed, unsigned, short, long, char, float, double.


\subsection{Assignments}
Binary assignments are taken care of in VisitBinAssign() inside MyProcessStmt class.


\subsection{Cascaded Assignments}
Cascaded assignments involving a chain of assignments (e.g. x = y = z = w; ) are taken care of in VisitBinAssign() inside MyProcessStmt class. \\

\noindent For cascaded assignments we have used two variables inside MyProcessStmt class:

\begin{itemize}
  \item isCascadedAssign: this boolean flag is set to true when we find RHS of assignment is again assignment in the AST of the statement.
  \item assignCount: an integer variable for number of pending assignments. Whenever we see an assignment operator in AST we increment assignCount to denote we have a pending (un-evaluated) assignment. When an assignment is evaluated, we decrease assignCount.
\end{itemize}

\noindent Cascaded assignments are evaluated from right to left. e.g. for statement x = y = z = w; when we reach to z = w visiting AST recursively, assignCount is 3 and isCascadedAssign is true. First z = w is evaluated and z is kept on expression stack, then y = z is evaluated and y is kept on expression stack, finally x = y is evaluated and assignCount becomes zero.


\subsection{Arithmetic Operators}
Arithmetic operators are +, -, *, / and \%. Table \ref{table:VisitBinAO} summarizes visit functions used for them inside MyProcessStmt class. They all use another function getLHSAndRHSForBO() to get left hand side and right hand side operand for their respective arithmetic binary operation.

\begin{table}[h!]
  \begin{center}
    \begin{tabular}{c | c | c}
    \hline\hline
    Arithmetic operation & Arithmetic operator & Visitor function \\
    \hline\hline
    addition & + & VisitBinAdd() \\
    subtraction & - & VisitBinSub() \\
    multiplication & * & VisitBinMul() \\
    division & / & VisitBinDiv() \\
    modulus & \% & VisitBinRem() \\
    \hline\hline
    \end{tabular}
    \caption{Arithmetic Operators}
    \label{table:VisitBinAO}
  \end{center}
\end{table}


\subsection{Compound Arithmetic Operators}
Compound arithmetic operators are +=, -=, *=, /= and \%=. Table \ref{table:VisitBinCAO} summarizes visit functions used for them inside MyProcessStmt class. They all use another function getLHSAndRHSForCAO() to get left hand side and right hand side operand for their respective compound arithmetic binary operation.

\begin{table}[h!]
  \begin{center}
    \begin{tabular}{c | c | c}
    \hline\hline
    Compound Arithmetic operation & Compound Arithmetic operator & Visitor function \\
    \hline\hline
    compound addition & += & VisitBinAddAssign() \\
    compound subtraction & -= & VisitBinSubAssign() \\
    compound multiplication & *= & VisitBinMulAssign() \\
    compound division & /= & VisitBinDivAssign() \\
    compound modulus & \%= & VisitBinRemAssign() \\
    \hline\hline
    \end{tabular}
    \caption{Compound Arithmetic Operators}
    \label{table:VisitBinCAO}
  \end{center}
\end{table}


\subsection{Relational Operators}
Relational operators are \textgreater, \textgreater=, \textless, \textless=, == and !=. Table \ref{table:VisitBinRel} summarizes visitor functions for them inside MyProcessStmt class. They all use another function getLHSAndRHSForRelBO() to get left hand side and right hand side operand for their respective relational binary operation.

\begin{table}[h!]
  \begin{center}
    \begin{tabular}{c | c | c}
    \hline\hline
    Relational operation & Relational operator & Visitor function \\
    \hline\hline
    greater than & \textgreater & VisitBinGT() \\
    greater than or equal to & \textgreater= & VisitBinGE() \\
    less than & \textless & VisitBinLT() \\
    less than or equal to  & \textless= & VisitBinLE() \\
    equality & == & VisitBinEQ() \\
    inequality & != & VisitBinNE() \\
    \hline\hline
    \end{tabular}
    \caption{Relational Operators}
    \label{table:VisitBinRel}
  \end{center}
\end{table}


\subsection{Unary Operators}
Unary operators are unary +, unary -, ++ (pre and post increment), -- (pre and post decrement) and ! (logical not). Table \ref{table:VisitUnary} summarizes visit functions used for them inside MyProcessStmt class for their respective unary operation.

\begin{table}[h!]
  \begin{center}
    \begin{tabular}{c | c | c}
    \hline\hline
    Unary operation & Unary operator & Visitor function \\
    \hline\hline
    unary plus & + & VisitUnaryPlus() \\
    unary minus & - & VisitUnaryMinus() \\
    pre-increment & ++ & VisitUnaryPreInc() \\
    post-increment & ++ & VisitUnaryPostInc() \\
    pre-decrement & - - & VisitUnaryPreDec() \\
    post-decrement & - - & VisitUnaryPostDec() \\
    logical not & ! & VisitUnaryLNot() \\
    \hline\hline
    \end{tabular}
    \caption{Unary Arithmetic Operators}
    \label{table:VisitUnary}
  \end{center}
\end{table}


\subsection{Bitwise Shift Operators}
Currently, bitwise shift operator are not supported fully as domain library for Interval, Octagon, Polyhedra does not support bitwise operations. Therefore when assignment statements involving bitwise shift operators are visited, abstract value for LHS variable of assignment is set to top, i.e. [$-\infty, +\infty$] in case of Interval. \\

\noindent Table \ref{table:VisitSh} summarizes visit functions used for bitwise shift operators inside MyProcessStmt class. These visitor functions use common functions getLHSAndRHSForShBO() (for \textless\textless, \textgreater\textgreater) and getLHSAndRHSForShCAO() (for \textless\textless=, \textgreater\textgreater=). 

\begin{table}[h!]
  \begin{center}
    \begin{tabular}{c | c | c}
    \hline\hline
    Bitwise shift operation & shift operator & Visitor function \\
    \hline\hline
    shift left & \textless\textless & VisitBinShl() \\
    shift right & \textgreater\textgreater & VisitBinShr() \\
    shift assign left & \textless\textless= & VisitBinShlAssign() \\
    shift assign right & \textgreater\textgreater= & VisitBinShrAssign() \\
    \hline\hline
    \end{tabular}
    \caption{Bitwise Shift Operators}
    \label{table:VisitSh}
  \end{center}
\end{table}

\subsection{Conditions}
Visit method for condition is VisitIfStmt() inside MyProcessStmt class. This supports special case of if (x) kind of statements as well. For if (x = y) statements (assignment in place of equality check), CIL transforms them to x = y; followed by if (x). Nested conditions are also supported.


\subsection{Loops}
Visit methods for while, do-while, for statements are VisitWhileStmt(), VisitDoStmt(), VisitForStmt() respectively inside MyProcessStmt class. Due to CIL transformations, all loops are transformed into while loops and VisitWhileStmt is called. This supports while (x) kind of statements as well. Other cases of conditions of while loop are simplified by CIL. \\

\noindent Nested loops are supported with CIL transformations. Break statements are transformed into a goto statement and a labelled null statement. Only CIL introduced forward goto statements are supported, no arbitrary jumps are supported.


\subsection{Implicit Cast}
C Analyzer's MyProcessStmt class takes care of all implicit casts using Clang API checks for statement class ImplicitCastExpr in case of all types of expressions. e.g. a) x = ui + 1; where x is int and ui is unsigned int variable. Using implicit cast, integer literal 1 is promoted to unsigned int. b) x = ui + y; where x, y are int and ui is unsigned int variable. Using implicit cast, y is promoted to unsigned int.


\subsection{C Style Explicit Cast}
\label{sec:CStyleExplicitCasts}

CIL introduces C style explicit cast during source-to-source transformation. Following C style explicit casts are supported inside VisitCStyleCastExpr() of MyProcessStmt class:

  \begin{itemize}

    \item IntegralCast: explicit cast to int, long, long long (both signed and unsigned), e.g. (int) x, (unsigned long) y
    \item FloatingToIntegral: cast to floating type assigning to integral type on LHS, e.g. int x; float a = 3.14; x = (double) a + 2.7182818281828;
    \item FloatingCast: explicit cast to float, double, long double, e.g. (float) 2.7182818281828, (double) w
    \item IntegralToFloating: cast to integral type assigning to real type on LHS, e.g. int x = 10; float a; a = (long) x + 6563565;

  \end{itemize}


\section{Assertion Verification}
Using invariants generated in different domains, we can verify some properties for C programs. There are two types of assertions added to verify these properties - implicit and explicit assertions.

\subsection{Implicit Assertions}
For implicit assertions, user need not specify any assertion or condition to be checked. C Analyzer code implicitly handles them. Currently, following implicit assertions are in place:

  \begin{itemize}
    \item Divide by zero: occurs when denominator for division is zero. For divide by zero, result is undefined, hence implicit assertion is thrown.

    \item Modulus zero: occurs when denominator for modulus operation is zero. For modulus zero, result is undefined, hence implicit assertion is thrown.

    \item Arithmetic overflow: occurs when an arithmetic operation (+, -, *, /, \%, ++, --) leads to overflow or underflow.

    \item Uninitialized variable used: occurs when a variable is used at program point $P$ but it is not initialized anywhere before $P$.

  \end{itemize}

\noindent Currently, analysis continues when any of above implicit assertion is violated, analysis does not terminate. e.g. when divide by zero is discovered, a warning/error message is sent and LHS of assignment will be set to top ($[-\infty, +\infty]$ in case of interval. 


\subsection{Explicit Assertions}

User defined assertions can be added to CIL transformed code using a dummy MYASSERT() function. We need not provide function body for MYASSERT() and assertion or condition is provided as argument to this dummy function. \\

e.g. MYASSERT(x \textgreater 0);\\

\noindent Advantage of implementing explicit assertions in this way is that Clang API will take care of assertion or condition passed as an argument to MYASSERT() while reading AST for expressions involving relational operators.


\pagebreak

\section{Adding New Domain}
\label{app:AddingNewDomain}

To add a new domain to current plug-and-play multiple domain architecture of C Analyzer, developer has to do following:

  \begin{itemize}

    \item Create following data structures: developer should create a structure to hold abstract value at any program point, a structure to store abstract value at the end of basic block, abstract value of positive of condition and negative of condition. Developer should manage its own internal structure (e.g. stack) for expressions and its own environment to keep track of variables in the program.

    \item Implement virtual functions: developer must create a class for new domain inheriting from Analyzer class in public mode and then override virtual functions defined in base class Analyzer (for reference see Analyzer.h and Interval.h).

    \item Add choice to use this new domain: developer must add choice for new domain in common interface inside MyASTVisitor, just like it is there for other domains (see MyASTVisitor.cpp). Create a class for new domain inheriting from Analyzer class in public mode. Create an object of class of new domain, pass arguments to parent class constructor and point Analyzer object to this domain object.

  \end{itemize}

\noindent Note:- If a domain chooses not to have a certain operation corresponding to virtual functions in Analyzer class (e.g. a domain may not use widening), even then developer should set abstract value to top or as the case may be for this domain in order to keep Analyzer interface common across domains.


\section{Code Features}

In its current implementation, C Analyzer code has following salient features:

  \begin{itemize}

    \item Code has been written in C++ nicely taking advantage of STL library wherever applicable and virtual polymorphism for plug-and-play multi-domain architecture.
    \item Abstract domains supported: It supports 4 abstract domains - Interval, Octagon, Polyhedra and Bit Vector.
    \item Extendibility: New abstract domains can be added and new AST visitor functions can be added for un-implemented features in C.
    \item Source location information is added for every expression and every statement. \\
    e.g.  x + y   at testdata/test\_labelStmt.c:13:10  (x + y starts at line no. 13, column no. 10)
    \item Building code: there is single makefile to build entire code base with all dependencies defined appropriately. All macros especially editable macros for user defined path settings are commented adequately.
    \item Documentation: C Analyzer code is nicely documented using Doxygen documentation tool. There are ample inline code comments and sufficient code debug statements throughout the code base.
    \item Classes and data structures: there are more than 10 classes and many data structures created.
    \item Functions: there are more than 230 functions written each mentioning input parameters and return type (only including common interface and Interval domain functions).
    \item Lines of Code: it has 14,000+ lines of code base (excluding code for Bit Vector domain).

  \end{itemize}


\section{Limitations}

Currently, following features in C are not supported by C Analyzer tool due to its primary focus on core C program features and plug and play multi-domain support:

  \begin{itemize}

    \item Arrays
    \item Structures and unions
    \item Pointers
    \item Dynamic memory allocation
    \item Function Calls
    \item Recursion
    \item Bitwise shift operators (limited support currently)
    \item Bitwise logical operators

  \end{itemize}


\pagebreak

\section{Concrete Example for Analysis}

This section provides examples of C programs being analyzed involving declarations, assignments, condition and loop for Interval abstract domain.

\subsection{Condition}
\begin{multicols}{2}
\begin{lstlisting}
Code:

int main()
{
    int x = 10;

    if (x > 0)   // B4
    {
        x = 100; // B3
    }
    else
    {
        x = -1;  // B2
    }

    return 0;    // B1
}
\end{lstlisting}
\hfill

\begin{lstlisting}
  abstract value after block terminator is processed
interval of dim (1,0):
       x in [10,10]
interval of dim (1,0):
       x in [1,+oo]
interval of dim (1,0):
       x in [-oo,0]

  @begin of block 3  abstract value after meet
interval of dim (1,0):
       x in [10,10]

  @begin of block 2  abstract value after meet
interval of dim (1,0): bottom

  @begin of block 1  abstract value after join
interval of dim (1,0):
       x in [100,100]

\end{lstlisting} 
\end{multicols}

\begin{figure}[h!]
  \centering
  \includegraphics[scale=0.45]{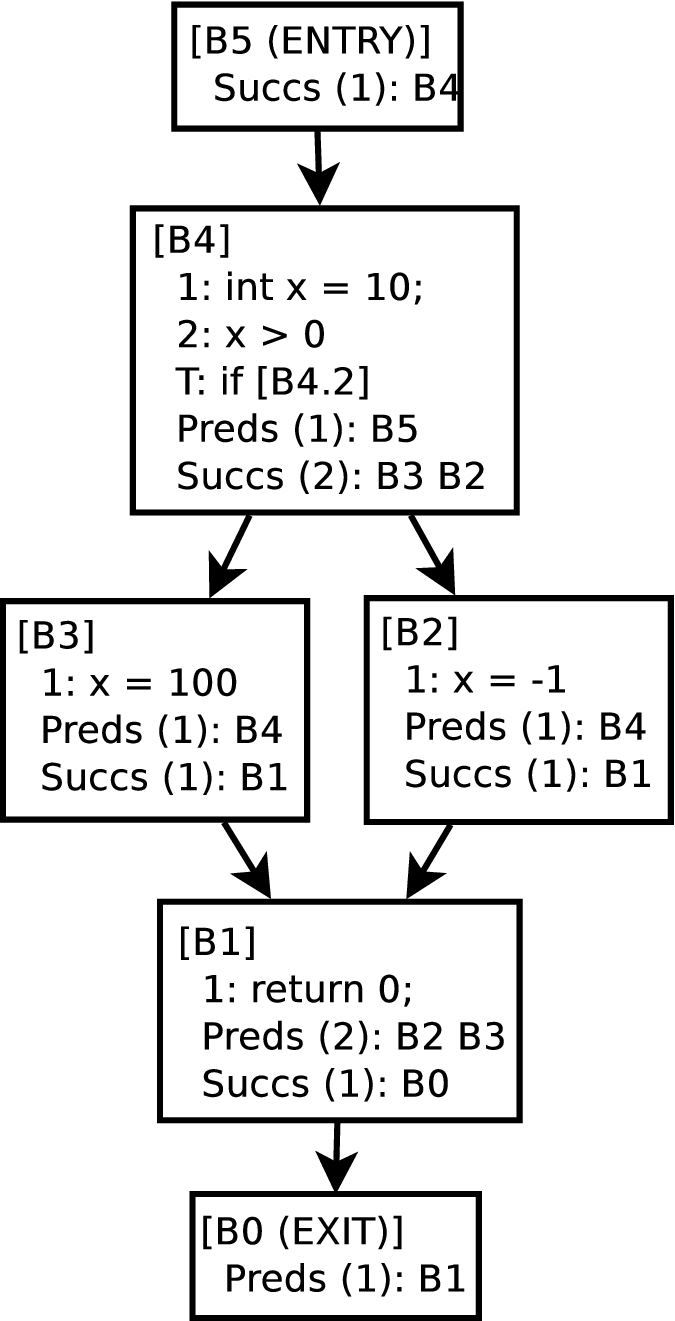}
  \caption{CFG to show meet and join}
  \label{fig:domains}
\end{figure}


\subsection{Loop}
Loop with Unrolling (NUM\_UNROLLINGS set to 5 times to delay widening):

\begin{multicols}{2}
\begin{lstlisting}
Code:

int main()
{
  int a = 6, b = 2;

  while(a>0)
  {
    a = a - 1;
  }

  b = a + b;

  return 0;
}
\end{lstlisting}
\hfill

\begin{lstlisting}
Fixed Point:
loopExitAbsValOld:
interval of dim (2,0):
       a in [0,5]
       b in [2,2]
loopExitAbsValCurrent:
interval of dim (2,0):
       a in [0,5]
       b in [2,2]
    
Resulting values:
abstract value:
interval of dim (2,0):
       a in [0,0]
       b in [2,2]
\end{lstlisting} 

\end{multicols}

\begin{figure}[h!]
  \centering
  \includegraphics[scale=0.5]{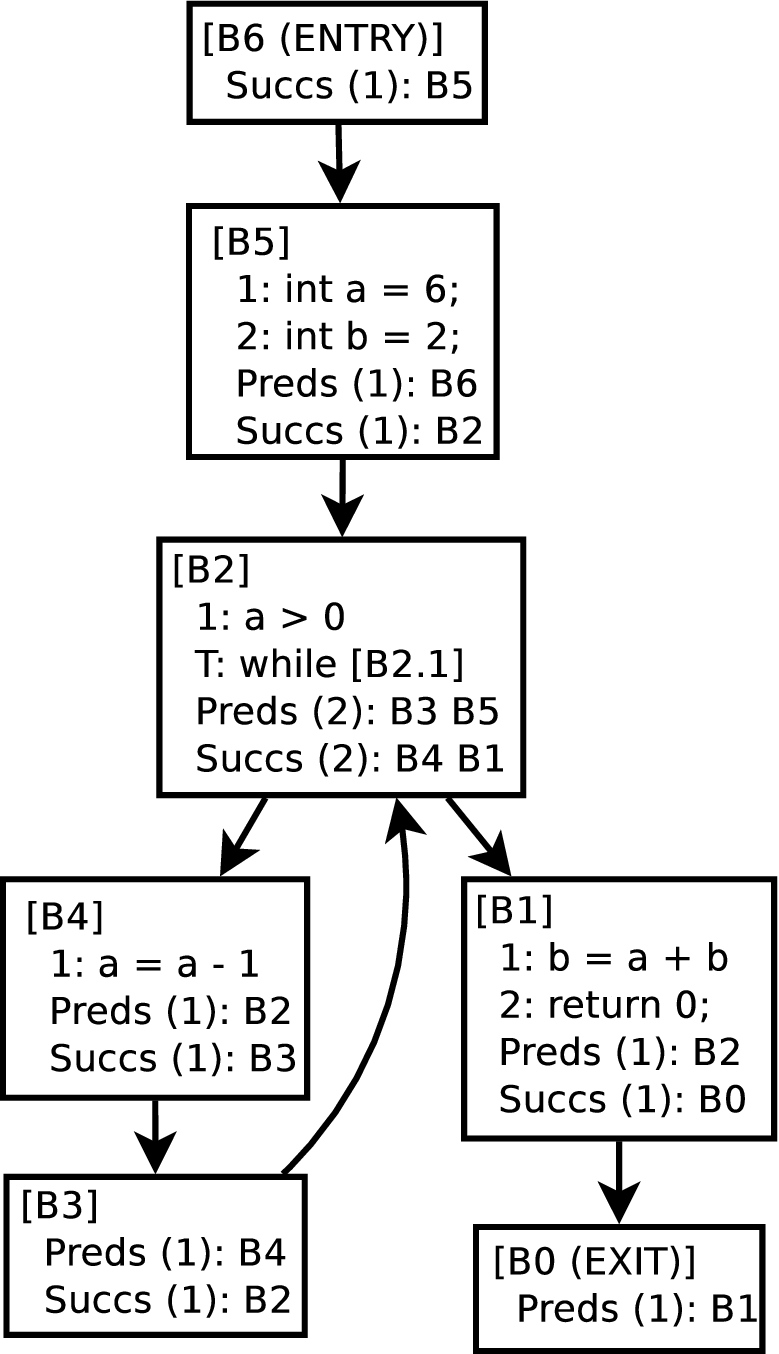}
  \caption{CFG to show widening}
  \label{fig:domains}
\end{figure}

\pagebreak

\noindent Loop without unrolling:

\begin{multicols}{2}

\begin{lstlisting}
Code:

int main()
{
  int a = 6, b = 2;

  while(a>0)
  {
    a =a-1;
  }

  b = a + b;

  return 0;
}
\end{lstlisting}
\hfill

\begin{lstlisting}
Fixed Point: 
loopExitAbsValOld:
interval of dim (2,0):
       a in [-oo,5]
       b in [2,2]
loopExitAbsValCurrent:
interval of dim (2,0):
       a in [-oo,5]
       b in [2,2]
    
Resulting values:
abstract value:
interval of dim (2,0):
       a in [-oo,0]
       b in [-oo,2]
\end{lstlisting} 

\end{multicols}


\chapter{Results and Discussions}
\label{sec:ResultsDiscussions}

C Analyzer provides a plug-and-play kind of architecture for multiple abstract domains to be used and new domains can be added easily by overriding appropriate virtual functions of Analyzer class. During pre-processing of CFG, this generates a list of basic blocks to be visited in order and a 2D vector to store edge information. Later, depending on abstract domain selected, this invokes appropriate implementation for the domain during analysis. \\

\noindent C Analyzer supports all types of declarations for integral and real data types, assignments, cascaded assignments, binary arithmetic operations, relational operations, unary arithmetic operations. Implicit cast and C style explicit cast expressions are also supported. C Analyzer also supports conditions (if-else, if-else ladder) and while loops. Nested loops can be analyzed. \\

\noindent Using source-to-source transformation tool CIL, do-while and for loops are simplified to while loops. Similarly, CIL transforms logical operators, switch statement and conditional operator into if-else ladder which is supported by C Analyzer. \\

\noindent For all above C constructs, using C Analyzer generated invariants for Interval, Octagon and Polyhedra, certain properties of programs can be analyzed. Implicit assertions such as divide by zero, modulus zero, integer overflow and uninitialized variable being used are checked. User defined assertions can be checked using dummy function MYASSERT(). Currently, it does not support arrays, structures, unions, pointers, function calls, recursion and dynamic memory allocation. \\

\noindent In future, C Analyzer tool can be evolved into a complete tool for scalable and more precise static analysis for large code base by leveraging more abstract domains to its disposal. Plug-and-play domain architecture can facilitate combining multiple abstract domains for more precise analysis and usage of abstract refinement techniques. \\

\noindent Features that are currently not supported (arrays, structures, unions, pointers, function calls, etc.), can be implemented in future. Bit Vector domain can be fully plugged into current plug-and-play architecture just like other domains.


\chapter{Summary and Conclusions}
\label{sec:SummaryConclusions}

This chapter summarizes this thesis on work done towards C Analyzer - a static program analysis tool built for C programs. In chapter \ref{sec:Introduction} Introduction, we emphasized the need for formal methods based techniques and got an overview of static program analysis and Abstract Interpretation. \\

\noindent In chapter \ref{sec:LiteratureSurvey} Literature Survey, we discussed some basic ideas and techniques to understand theory of Abstract Interpretation better. We discussed some relevant topics under lattice theory, abstract domains, Galois connection, collecting semantics, abstract operators, etc. \\

\noindent Chapter \ref{sec:CAnalyzer} C Analyzer Implementation, provided motivation for plug-and-play domain architecture and implementation specific details on C Analyzer beginning with high level overview of the tool. We discussed CIL transformations, data structures and classes of C Analyzer, general processes and algorithms used during analysis - CFG generation, CFG iteration, computation of abstract summary. Later, we discussed features implemented, code features and limitations of the tool. \\

\noindent Chapter \ref{sec:ResultsDiscussions} Results and Discussions, summarized major contributions of this project work and a road map for future work to be done in this direction. Next chapter Bibliography gives references used. \\

\noindent C Analyzer tool can evolve into a complete tool for scalable static analysis for large code base and can leverage more abstract domains to its disposal. Features that are currently not supported, can be implemented in due course for better coverage of C constructs.







\begin{thebibliography}{10}

\addcontentsline{toc}{chapter}{Bibliography}

\bibitem{WIKI}Wikipedia for several conceptual articles, especially under chapter Literature Survey \url{http://en.wikipedia.org/}

\bibitem{Intro-AI}Patrick Cousot, Introduction to Abstract Interpretation \\
\url{http://www.di.ens.fr/~cousot/AI/IntroAbsInt.html}.

\bibitem{AI-FW}Patrick Cousot \& Radhia Cousot. Abstract interpretation frameworks. Journal of Logic and Computation, 2(4):511—547, August 1992.

\bibitem{Intro-ExAI}Examples of abstract-interpretation-based static analysis \\
\url{http://www.di.ens.fr/~cousot/AI/#tth\_sEc4}.

\bibitem{WIDNARO}Agostino Cortesi and Matteo Zanioli: Widening and Narrowing Operators for Abstract Interpretation, Computer Languages, Systems \& Structures 37(1): 24-42 (2011)

\bibitem{AI-SUTRE}Gregoire Sutre 2008, slides on Software Verification \\
\url{www.mpi-inf.mpg.de/vtsa08/slides/sutre1.pdf} and \\
\url{www.mpi-inf.mpg.de/vtsa08/slides/sutre2.pdf}

\bibitem{ABS-INT-TUT}Tutorial on Abstract Interpretation: \burl {https://ti.arc.nasa.gov/m/tech/rse/publications/papers/cglobalsurveyor/abs\_int\_tutorial.ppt}



\bibitem{Intro-bugs}Software Bugs \url{http://www5.in.tum.de/~huckle/bugse.html}.

\bibitem{Intro-SPA}Static Program Analysis, \url{http://www.irisa.fr/lande/jensen/spa.html}


\bibitem{LIT-Mine}Antoine Min$\acute{e}$: 2006, \textsl{The Octagon Abstract Domain}.

\bibitem{Ctree}Ctree - an implementation of AST generation using Flex/Bison based parser \\
\url{http://sourceforge.net/projects/ctool/files/ctree/}.

\bibitem{Ctree-Lex}ANSI C Grammar - Lex \url{http://www.lysator.liu.se/c/ANSI-C-grammar-l.html}

\bibitem{Ctree-Yacc}ANSI C Grammar - Yacc \url{http://www.lysator.liu.se/c/ANSI-C-grammar-y.html}

\bibitem{APRON}APRON \url{http://apron.cri.ensmp.fr/library/}

\bibitem{LLVM}The LLVM Compiler Infrastructure Umbrella Project \url{http://llvm.org/}.

\bibitem{LLVM-Clang}Clang - C, C++, Objective-C compiler \url{http://clang.llvm.org/}.

\bibitem{LLVM-CSAC}Clang - Static Analyzer Checker User Guide \\
\url{http://clang-analyzer.llvm.org/checker\_dev\_manual.html}.

\bibitem{LLVM-LV}Chris Lattner and Vikram Adve. LLVM: A Compilation Framework for Lifelong Program Analysis \& Transformation. In CGO'04: \textsl{Proceedings of the International Symposium on Code Generation and Optimization: Feedback-directed and Run-time Optimization, 2004.}

\bibitem{LLVM-aosa}LLVM: The Architecture of Open Source Applications Vol-I chapter-11 \\
\url{http://www.aosabook.org/en/llvm.html}.

\bibitem{LLVM-REPO}Clang/LLVM Maturity Report, Dominic Fandrey Proceedings of the Summer 2010 Research Seminar, Computer Science Dept., University of Applied Sciences Karlsruhe, June 2010.

\bibitem{StmtRef}Clang's Stmt class reference online at LLVM website: \\
\url{http://clang.llvm.org/doxygen/classclang\_1\_1Stmt.html}

\bibitem{RecASTV}Clang's RecursiveASTVisitor class reference: \\
\url{http://clang.llvm.org/doxygen/classclang\_1\_1RecursiveASTVisitor.html}

\end{thebibliography}
\end{document}